\documentstyle[12pt]{article}
\begin{document}                             
\title{CP Violating Rate
Difference Relations for $B\to PP$
and $B \to PV$ in Broken SU(3)}
\author{M. A. Dariescu$^a$\thanks{On leave of
absence from Department of Theoretical Physics,
{\it Al. I. Cuza} University, Bd. Copou no. 11,
6600 Ia\c{s}i, Romania}, N. G. Deshpande$^a$,
X.-G. He$^b$ and G. Valencia$^c$}
\date{}
\maketitle

\begin{center}
$a$ Institute of Theoretical Science,
University of Oregon, OR 97403, USA\\
$b$ Department of Physics, National Taiwan University,
Taipei, Taiwan\\ 
$c$ Department of Physics, Iowa State University, Ames, Iowa 50011, USA
\end{center}

\begin{abstract}

Within the standard model there exist certain relations 
between CP violating rate differences in $B$ decays in the 
$SU(3)$ limit. We study $SU(3)$ breaking 
corrections to these relations in the case of charmless, 
hadronic, two body $B$ decays using the improved factorization 
model of Ref.\cite{3}. We consider the cases 
$B \to PP$ and $B \to PV$ for both $B_d$ and $B_s$ mesons. 
We present an estimate for $A_{CP}(\pi^- \pi^+)$
in terms of $A_{CP}(K^- \pi^+)$.

\end{abstract}

$B$ decays are a subject of very active research
at present since they provide
useful information on the dynamics of strong and electroweak
interactions for testing the Standard Model 
(SM) and models beyond and are ideally suited for
a critical analysis of CP violation. The 
mixing induced CP asymmetry in $\bar{B}^0 \to \psi K_s$
versus $B^0 \to \psi K_s$ has already provided an
accurate measurement of $\sin 2 \beta$ \cite{1,2}.
This result is in excellent agreement with the SM.
Other mixing studies, such as $\bar{B} \to \pi^- \pi ^+$, are
underway for determining $ \sin 2 \alpha$, but require more data
to reduce the  theoretical uncertainty 
associated with penguin contributions.
Rate asymmetry measurements in the branching ratios
of $B$ mesons into mesons involving
light quarks are also underway.
These shed light on direct CP violation in the amplitudes.
Analyses of these decays to extract fundamental parameters of the
SM are more difficult because of theoretical
uncertainty in the calculation of hadronic modes. 
In general, these asymmetries arise from interference
of a Cabbibo suppressed tree amplitude with
a (possibly enhanced) penguin amplitude. 
As such, these asymmetries are sensitive to
contributions through loops, that could involve physics
beyond the SM. Thus, the study of direct CP violation can 
be a powerful tool to probe for physics beyond the SM  
if the theoretical uncertainty can be reduced.

The goal of this letter is to study the direct CP violation 
asymmetry in a class of processes where there has been recent 
theoretical progress. These processes involve $B$
decays into two light pseudoscalars $P_1 P_2$ or into a
light pseudoscalar and a light vector meson $PV$. 
We identify relations between rate asymmetries which
are valid in the SU(3) limit in the standard model, and we 
compute SU(3) breaking corrections to them using the QCD 
improved factorization model of Ref.\cite{3}. We also discuss 
additional relations which are valid in the SU(3) limit when 
annihilation contributions are neglected. 

CP violation in the SM arises solely from
the phase in the $3 \times 3$ unitary CKM matrix, $V_{CKM} = (V_{ij})$,
and any CP violating observable is proportional to 
$Im(V_{ij}V_{il}^*V_{kj}^*V_{kl})$,
with $i\neq k$ and $j\neq l$.
This simple property has important implications, as
for example it leads to relations among CP violating 
rate differences, $\Delta^B_{PP} = \Gamma(B\to PP) - 
\bar \Gamma(\bar B \to \bar P\bar P)$, for different decay modes. 
For instance, it has been shown that,
with SU(3) flavor symmetry, 
when small annihilation contributions
and phase space differences are neglected,  
naive factorization yields the relation\cite{4}
\begin{eqnarray}
\Delta^{\bar B^0}_{\pi^- \pi^+} \approx -
{f^2_\pi\over f^2_K} \Delta^{\bar B^0}_{K^- \pi^+} \, .
\label{relg}
\end{eqnarray}
This can be used to test the SM  CP violation, or
to predict one rate difference if
the other one is known. The above equation leads to a
relation for the CP violating rate asymmetry,
\begin{equation}
A_{CP}(\pi^-\pi^+) \approx - {f_\pi^2 \over f^2_K} 
{Br(K^- \pi^+)\over Br(\pi^-\pi^+)} \, A_{CP}(K^- \pi^+),
\label{rela}
\end{equation}
where $Br(PP)$ are the CP averaged branching ratios, 
$Br(\pi^-\pi^+) = (5.2 \pm 0.6)\times 10^{-6}$
and $Br(K^- \pi^+) = (18.6\pm 1.1)\times 10^{-6}$\cite{1,2}. 
Eq.~\ref{rela} implies the following relation between
the corresponding CP asymmetries: $A_{CP}(\pi^-\pi^+) 
\approx - 2.4 \, A_{CP} (K^- \pi^+)$.

Preliminary data on these asymmetries is just
emerging, with BaBar reporting $A_{CP}(\pi^- \pi^+) = - 0.30
\pm 0.25 \pm 0.04$, $A_{CP}(K^- \pi^+) =
-0.102 \pm 0.050 \pm 0.016$\cite{1} and Belle reporting 
$A_{CP}(\pi^- \pi^+ ) = 0.94^{+0.25}_{-0.31}
\pm 0.09$\cite{2}. At the moment the two experiments disagree on 
the value of  $A_{CP}(\pi^- \pi^+)$ but they still have very large errors. 

The most important question for theory is to establish the 
precision of Eqs.~\ref{relg} and \ref{rela} within the standard model, or equivalently 
to estimate the corrections they receive. One can easily identify two 
sources of corrections for Eqs.~\ref{relg} and \ref{rela}: 
annihilation contributions, and SU(3)breaking effects. 
Even though the relation Eq.~\ref{relg} already includes some SU(3)
breaking effects in the factor $f^2_\pi/f^2_K$, it is necessary to 
have better control over these corrections 
in order to test the Standard Model.

To begin our analysis of the $B \to PP$ modes, we 
first note that there are several relations 
among the rate differences in these decays that follow 
from SU(3) flavor symmetry in the SM. There are also other 
relations such as Eq.~\ref{relg} which rely both on SU(3) 
symmetry and on the neglect of annihilation amplitudes. 
It is easy to understand the origin of these relations. 
The decay amplitude for $B \to PP$ can be parameterized as
\begin{eqnarray}
A(B\to PP) = V_{ub}V_{uq}^* T^B_{PP} 
+ V_{cb}V_{cq}^* P^B_{PP},
\end{eqnarray}
and can be decomposed into SU(3) invariant amplitudes 
according to the representation of the
effective Hamiltonian \cite{4}.
SU(3) symmetry predicts that the amplitudes for 
$\bar B^0 \to \pi^- \pi^+, K^- \pi^+$
and $\bar B_s^0 \to \pi^- K^+, K^- K^+$
are related and this can be proved by writing the
decay amplitudes in terms of the SU(3)
invariant amplitudes as
\begin{eqnarray}
&&
T^{\bar B^0}_{\pi^- \pi^+} = T^{\bar B_s^0}_{K^- K^+} = 
2 A^T_{\bar 3} + C^T_{\bar 3} + C^T_6 
+ A^T_{\overline{15}} + 3 C^T_{\overline{15}},
\nonumber\\
&&
T^{\bar B^0}_{K^- \pi^+} = T^{\bar B_s^0}_{\pi^- K^+} = 
C^T_{\bar 3} + C^T_6 
- A^T_{\overline{15}} + 3 C^T_{\overline{15}},
\end{eqnarray}
where $A_i$ indicate the annihilation contributions.  
Both model calculations\cite{3}, and fits to experimental data\cite{he}
indicate that these annihilation amplitudes are small. 
The penguin amplitudes $P^B_{PP}$ can
be parameterized in a similar way.

We note that, even though $T(P)^{\bar B^0}_{\pi^-\pi^+} =
T(P)^{\bar B_s^0}_{K^-K^+}$ and
$T(P)^{\bar B^0}_{K^- \pi^+} = 
T(P)^{\bar B_s^0}_{\pi^- K^+}$ in 
the SU(3) limit, there are no simple relations between
the branching ratios for these decays, because the 
CKM factors in the $T$ and $P$ amplitudes are different. 
However, because $\Delta^B_{PP} \sim Im(TP^*)
Im(V_{ub}V_{uq}^*V_{cb}^*V_{cq})$ 
and $Im(V_{ub}V_{ud}^*V_{cb}^*V_{cd}) = 
 - Im(V_{ub}V_{us}^*V_{cb}^*V_{cs})$ from the 
unitarity  of the CKM matrix,  
we have the following relations among
the CP violating rate differences:
\begin{eqnarray}
\Delta^{\bar B^0}_{\pi^-\pi^+} = - 
\Delta^{\bar B_s^0}_{K^- K^+},\;\;\;\;
\Delta^{\bar B^0_s}_{\pi^- K^+} = - 
\Delta^{\bar B^0}_{K^- \pi^+} \, .
\label{relsu3}
\end{eqnarray}
These relations can be obtained by interchanging the
$d$ and $s$ quarks (U-spin symmetry).  
If annihilation contributions are neglected,
all the amplitudes $T^{\bar B^0}_{\pi^-\pi^+}$,
$T^{\bar B^0}_{K^- \pi^+}$, 
$T^{\bar B_s^0}_{\pi^- K^+}$
and $T^{\bar B_s^0}_{K^- K^+}$ are 
approximately equal and one gets additional relations, 
\begin{eqnarray}
\Delta^{\bar B^0}_{\pi^-\pi^+} = 
-\Delta^{\bar B_s^0}_{K^- K^+} 
= \Delta^{\bar B_s^0}_{\pi^- K^+} = 
-\Delta^{\bar B^0}_{K^- \pi^+}.
\label{relsc}
\end{eqnarray}

Similar relations exist as well for decays with 
neutral mesons in the final state. These relations are more 
complicated than those of Eq.~\ref{relsc} because there are 
more $d$ and $s$ quarks to interchange, and consequently it is 
harder to study the effect of SU(3) breaking in that case.  
For the remainder of this paper we concentrate 
on the relations in Eqs.\ref{relsu3} and \ref{relsc}.

One must be careful, however, about
the validity of these relations.
In the exact SU(3) limit, the $\Delta^B_{PP}$
are also exactly zero because the standard model 
conserves CP when $m_s=m_d$. It is well known that 
in this case it is possible to remove the phase in 
the CKM matrix with an appropriate rotation
among $d$ and $s$ quarks. In order to have a non-zero 
$\Delta^B_{PP}$, one can not have an exact SU(3) symmetry. 
In order to have CP violation in the 
standard model no two quarks with the same charge 
can have the same mass; as long as $m_s \neq m_d$ there 
will be CP violation regardless of how small these masses are. 
The relations in Eq.~\ref{relsu3} are thus valid and 
non-trivial in the limit where $(m_s-m_d)$ is much smaller 
than the QCD scale, but not zero.

When SU(3) breaking effects are included,
the above mentioned relations will be modified and one 
needs a good understanding of these effects before using 
the relations to test the standard model. Our limited understanding 
of the strong interaction dynamics at low energies makes this task 
quite difficult. In what follows we illustrate the SU(3) breaking 
corrections that arise within the QCD improved factorization 
model of Ref.~\cite{3}.

Within this approach, the relevant decay amplitudes
for $ B \to PP$ are given by\cite{3,5}
\begin{eqnarray}
A(\bar B^0 \to \pi^- \pi^+) 
&=& i {G_F \over \sqrt{2}} (m^2_B - m^2_\pi) 
F^{B\to \pi}_0(m^2_\pi) 
f_\pi [V_{ub}V_{ud}^* a_1(\pi\pi)
\nonumber \\*
&+& V_{pb}V_{pd}^*(a^p_4(\pi\pi)+ a^p_{10}(\pi\pi) 
+ r^\pi_\chi (a_6^p(\pi\pi) + a_8^p(\pi\pi)))]
\nonumber \\*
&+& i{G_F\over \sqrt{2}} f_B f^2_\pi [
V_{ub}V_{ud}^* b_1(\pi\pi)
\nonumber\\*
&+& (V_{ub}V_{ud}^* + V_{cb}V_{cd}^*)(b_3(\pi\pi) 
+ 2b_4(\pi\pi) 
\nonumber \\*
&-& {1\over 2} (b_3^{EW}(\pi\pi) - 
b_4^{EW}(\pi\pi))]
\nonumber\\ 
A(\bar B^0 \to K^- \pi^+) 
&=& i {G_F\over \sqrt{2}} (m^2_B - m^2_\pi) 
F^{B\to \pi}_0(m^2_K)
f_K [V_{ub}V_{us}^*a_1(K \pi )
\nonumber\\*
&+& V_{pb}V_{ps}^*(a^p_4(K \pi )+ a^p_{10}(K \pi) 
+ r^K_\chi (a_6^p(K \pi) + a_8^p(K \pi)))] 
\nonumber\\*
&+& i{G_F\over \sqrt{2}} f_B f_\pi f_K [
(V_{ub}V_{us}^*+ V_{cb}V_{cs}^*)(b_3(K \pi ) 
 - {1\over 2} b_3^{EW}(K \pi)], 
\nonumber \\*
\label{abppimpfac}
\end{eqnarray}
where $p$ is summed over $u$ and $c$, $r^\pi_\chi = 
2m^2_\pi/m_b(m_u+m_d)$,
$r^K_\chi = 2m^2_K/m_b(m_u + m_s)$ and
\begin{eqnarray}
a_1(M_1 M_2) & = & c_1 + {c_2 \over N_c}
\left[ 1 + {C_F \alpha_s \over 4 \pi} 
(V_{M_1} +  {4\pi^2\over N_c} H_{M_1 M_2}) \right],
\nonumber\\  
a^p_4(M_1 M_2) & = &
c_4 + {c_3\over N_c} \left[ 1 + {C_F \alpha_s \over 4 \pi} 
(V_{M_1} + {4\pi^2\over N_c} H_{M_1 M_2}) \right] + 
{C_F \alpha_s \over 4\pi N_c} P^p_{M_1,2}, 
\nonumber\\ 
a^p_6(M_1 M_2) & = & c_6 + {c_5\over N_c} 
\left( 1 - 6 {C_F \alpha_s \over 4 \pi} \right) + 
{C_F\alpha_s \over 4\pi N_c} P^{p}_{M_1, 3},
\nonumber\\ 
a^p_{8}(M_1 M_2) & = & c_8 + {c_7 \over N_c} 
\left( 1 - 6{C_F \alpha_s\over 4 \pi} \right) 
+ {\alpha\over 9\pi N_c} P^{p,EW}_{M_1, 3},
\nonumber\\ 
a^p_{10}(M_1 M_2) & = & c_{10} + {c_9\over N_c}
\left[ 1 + {C_F \alpha_s \over 4 \pi} (V_{M_1} + 
{4\pi^2\over N_c} H_{M_1 M_2}) \right] + 
{\alpha \over 9 \pi N_c} P^{p,EW}_{M_1, 2}, 
\\
b_1(M_1M_2) & = & {C_F \over N_c^2} c_1 A_1^i(M_1M_2),
\nonumber \\
b_3(M_1M_2) & = & {C_F \over N_c^2} [c_3 A_1^i(M_1M_2) + 
c_5 (A^i_3(M_1M_2) + A_3^f(M_1M_2))
\nonumber \\*
& + & N_c c_6 A^f_3(M_1M_2)],
\nonumber\\
b_4(M_1M_2) & = & {C_F\over N_c^2} [c_4 A_1^i(M_1M_2) + 
c_6 A^i_2(M_1M_2)],
\nonumber\\
b_3^{EW}(M_1M_2) & = & {C_F\over N_c^2} [c_9 A_1^i(M_1M_2) + 
c_7 (A^i_3(M_1M_2) + A^f_3(M_1M_2)) 
\nonumber \\*
& + & N_C C_8 A_3^f(M_1M_2)], 
\nonumber\\
b_4^{EW}(M_1M_2) & = & {C_F\over N_c^2} [c_{10} A_1^i(M_1M_2)
+ c_8 A^i_2(M_1M_2)],
\label{facsbpp}
\end{eqnarray}
where $C_F = (N^2_c-1)/2N_c$ and $N_c = 3$
is the number of colors and $c_i$ are the Wilson
coefficients.
The vertex, the hard gluon exchange with the
spectator, and the penguin contributions are:
\begin{eqnarray}
V_M &=& 12 \ln {m_b\over \mu} - 18 + \int^1_0 dx g(x) 
\Phi_M(x),
\nonumber\\
P^p_{M,2} &=&c_1 \left[ {4\over 3} \ln{m_b\over \mu} + 
{2\over 3} - G_M(s_p) \right]
\nonumber \\*
& + & c_3 \left[{8\over 3}\ln{m_b\over \mu} + 
{4\over 3} - G_M(0) - G_M(1) \right]
\nonumber\\*
&+& (c_4+c_6) \left[ {4n_f\over 3}\ln{m_b\over \mu} - 
(n_f-2) G_M(0) -G_M (s_c) -G_M(1) \right]
\nonumber\\*
& -& 2 c_{8g}^{eff}\int^1_0 {dx\over 1-x}\Phi_M(x),
\nonumber\\
P^{p,EW}_{M,2}& =& (c_1 + N_c c_2) \left[ {4\over 3} 
\ln{m_b\over \mu} + {2\over 3} - G_M(s_p) \right] - 
3 c_{7\gamma}^{eff} \int^1_0 {dx\over 1-x} \Phi_M(x),
\nonumber\\
P^p_{M,3} &=&c_1 \left[ {4\over 3} \ln{m_b\over \mu} 
+ {2\over 3} - \hat G_M(s_p) \right]
\nonumber \\*
& + & c_3 \left[{8\over 3}\ln{m_b\over \mu} + 
{4\over 3} - \hat G_M(0) - \hat G_M(1) \right]
\nonumber\\*
&+&(c_4+c_6) \left[ {4n_f\over 3}\ln{m_b\over \mu} - 
(n_f-2) \hat G_M(0) -\hat G_M(s_c) -\hat G_M(1) \right] 
\nonumber \\*
&-& 2 c_{8g}^{eff},
\nonumber\\
P^{p,EW}_{M,3} &=& (c_1 + N_c c_2) 
\left[{4\over 3} \ln{m_b\over \mu} + 
{2\over 3} - \hat G_K(s_p) \right] - 3 c_{7\gamma}^{eff},
\nonumber\\
H_{M_1 M_2} &=& {f_B f_\pi\over m_B \lambda_B 
F^{B\to \pi}_0(0)} \{ \int^1_0{dx\over 1-x}
\Phi_{M_1}(x) \int^1_0{dy\over 1-y} \Phi_{M_2}(y) 
\nonumber\\
&+& {\alpha_s(\mu_s) r^\pi_\chi(\mu_s) \over 
\alpha_s(\mu_h)}\int^1_0{dx\over x}
\Phi_{M_1}(x) \int^1_0 {dy\over 1-y}\Phi_p(y) \},
\nonumber \\*
A^i_j (M_1M_2) &=& \pi \alpha_s \int^1_0 dx dy \,
F_j^i(x,y) \, , \; j= \overline{1,3} \, ,
\nonumber \\*
A^f_3(M_1M_2) &=& \pi \alpha_s \int^1_0 dx dy \,
F_3^f(x,y) \, ,
\label{ffexp}
\end{eqnarray}
with
\begin{eqnarray}
F_1^i (x,y) & = &
\left \{ \Phi_{M_1}(x) \Phi_{M_2}(y)
\left[ {1\over y(1-x \bar y)} + 
{1\over y \bar x^2} \right] + 
{4 \mu_{M_2} \mu_{M_1}\over m^2_b} 
{2\over \bar x y} \right \},
\nonumber\\
F^i_2(x,y) &=& 
\left \{ \Phi_{M_1}(x) \Phi_{M_2}(y)
\left[ {1\over \bar x(1-x\bar y)} 
+ {1\over y^2 \bar x} \right] + 
{4 \mu_{M_2} \mu_{M_1}\over m^2_b} 
{2\over \bar x y} \right\},
\nonumber\\
F^i_3(x,y) &=& 
\left \{{2\mu_{M_2}\over m_b} 
\Phi_{M_1}(x){2\bar y\over \bar x y(1-x\bar y)} 
- {2\mu_{M_1}\over m_b} \Phi_{M_2}(y) 
{2x\over \bar x y(1-x\bar y)} \right \},
\nonumber\\
F^f_3(x,y) &=& 
\left \{{2\mu_{M_2}\over m_b} 
\Phi_{M_1}(x){2(1+\bar x)
\over \bar x^2 y} + {2\mu_{M_1}\over m_b}
\Phi_{M_2}(y) {2(1+y)\over \bar x y^2} \right\},
\label{fdist}
\end{eqnarray}
where $\bar x= 1-x$, $\bar y = 1-y$
and the parameter $2\mu_M/m_b$ coincides with $r_\chi$.
The functions $g(x)$, $G_M(x)$
and $\hat G_M(x)$ are given by \cite{3}
\begin{eqnarray}
g(x) & = & 3 \left( {1-2x\over 1-x}\ln x - i\pi \right) 
\nonumber \\*
& + &
\left[ 2 {\rm Li}_2(x) -\ln^2x + {2\ln x\over 1-x} - 
(3+2i\pi)\ln x -(x \to 1-x) \right],
\nonumber\\
G(s,x) & = & - \, 4 \int^1_0 du \, u(1-u) \ln[s-u(1-u)x],
\nonumber \\
G_M(s) & = & \int^1_0 dx \, 
G(s-i\epsilon, 1-x) \Phi_M(x), 
\nonumber\\
\hat G_M(s) & = & \int^1_0 dx \, 
G(s-i\epsilon, 1-x) \Phi_p(x),
\label{ffun}
\end{eqnarray}
where $ s_i  = m_i^2/m_b^2$ are the mass ratios
for the quarks involved in the penguin diagrams,
while $\Phi_M(x)$ and $\Phi_p(x)$ are the
distribution amplitudes of the $M$ meson.
The twist-3 distribution amplitude, $\Phi_p(x)$, is
equal to 1, to the order 
considered in the calculation. 
The distribution amplitude $\Phi_M(x)$ has
the following expansion in Gegenbauer polynomials\cite{3,6}
\begin{eqnarray}
\Phi_M(x) = 6x(1-x)[ 1+ \alpha_1 C^{(3/2)}_1(2x-1) + 
\alpha_2C^{3/2}_2(2x-1) + ...],
\label{ge}
\end{eqnarray}
with $C^{3/2}_1(u) = 3 u$ and
$C^{3/2}_2(u) = (3/2)(5u^2-1)$,
and is different for $\pi$ and $K$.
For $\pi$, the distribution in $x$
must be even because the $u$ and $d$ quarks have negligible masses 
and their
distributions inside the pion are symmetric.  
This dictates $\alpha_1^\pi = 0$. The coefficient
$\alpha^\pi_2$ is estimated to be $0.1 \pm 0.3$. 
For K, the $u$ (or $d$) and $s$ quarks
inside the kaon are different, leading to
an asymmetry in the $x$ distribution.
So a non-zero value for $\alpha^K_1$ is needed
and it is estimated to be
$0.3 \pm 0.3$, while $\alpha^K_2 = 0.1 \pm 0.3$\cite{3,6}. 

One also has to consider divergences contained in
the hard scattering and annihilation contributions.
The divergent part in the hard scattering comes from
$X_H = \int^1_0 \Phi_p(x) dy/(1-y) \approx 
\int^1_0 dy/(1-y)$, while the divergent part in
the annihilation is of the same form at leading order. 
These divergences are logarithmic and, 
in principle, would be absent in a full theory. 
Here, we follow Ref.\cite{3} to introduce an infrared cut-off at
$\Lambda_h = 0.5$ GeV and use 
\begin{equation}
X_{H(A)} = \, \ln {m_B \over \Lambda_h}\,.
\label{ircut}
\end{equation}
The final results are insensitive
to the precise value of the cut-off.
As for the numerical inputs,
we will use the values of the Wilson
coefficients at $\mu = m_b$\cite{3}, 
$\mu_h= \sqrt{\Lambda_h \mu}$,
$\Lambda_h = 0.5 $ GeV, $\lambda_B = 0.350$ GeV. 

The decay amplitudes for $\bar B_s^0 \to \pi^- K^+$ and 
$\bar B_s^0 \to K^- K^+$ can be obtained
from Eq.~\ref{abppimpfac}, by using the appropriate transition
form factor $F^{B_s\to K}_0$ and 
by changing $1/m^2_B\lambda_B$ to
$1/m^2_{B_s} \lambda_{B_s}$ in $H_{M_1M_2}$.

Putting everything together, we are now able to
estimate the size of different contributions
and, as expected, we find that the annihilation contributions are 
small.

Eq.~\ref{relg} incorporates SU(3) breaking effects only 
through the difference in the decay constants $f_\pi$ and 
$f_K$ as they appear in naive factorization. To improve 
on this approximation we need to consider other sources 
of SU(3) breaking. For example there are 
SU(3) breaking mass differences in both the initial $B$ mesons
and in the final state mesons. 
These mass differences induce corrections that are proportional to
$m^2_M/m_B^2$ and are therefore small.
The decay amplitudes are proportional
to the decay constant $f_{M_1}$ of $M_1$
and to the transition form factor $F^{B\to M_2}_0$,
depending on which $B$ is decaying 
into which final state. These form factors can also introduce
SU(3) breaking effects. 
For $\bar B^0\to \pi^-\pi^+$ and
$\bar B^0\to K^- \pi^+ $, all the corrections 
mentioned above account for the factor
$f^2_\pi/f^2_K$ in Eq.~\ref{relg}. 
Additional SU(3) breaking effects can arise from the
distribution amplitude $\Phi_M(x)$,
which is different for the $d$ and $s$ quarks.
The important effect arises from the twist two
distribution amplitudes. In our calculation we 
have used a constant $\Phi_p = 1$
for the twist-3 distribution amplitude so we have 
neglected SU(3) breaking in this term. However, SU(3) 
breaking effects in this term should also be taken into 
consideration at higher order. 

To summarize, the SU(3) 
breaking effects that we do include are the difference 
in the decay constants and form factors; and the 
difference in the $\alpha_1$ and $\alpha_2$ terms 
that appear in the twist-2 distribution amplitude. 
With these effects taken into account, 
and using $s_u=s_d=s_s=0$ and $s_c=(1.3/4.2)^2$
in Eq.~\ref{ffexp}, the relation in Eq.~\ref{relg} turns into
\begin{equation}
{\Delta^{\bar B^0}_{\pi^-\pi^+}\over 
\Delta^{\bar B^0}_{K^- \pi^+}}
\approx -{f^2_\pi\over f^2_K} 
\left[ \frac{1- 0.748 \alpha^\pi_1 - 0.109 
\alpha^\pi_2 - 0.017 H_{\pi \pi} -0.004\delta_A^\pi }{ 
1- 0.748 \alpha^K_1  - 0.109 
\alpha^K_2 - 0.017 H_{K \pi} + 0.0004 \delta_A^K} \right],
\label{relgmod}
\end{equation}
where $\delta_A^\pi = 1-1.34 X_A^\pi - 0.36 (X_A^\pi)^2$ and $\delta_A^K 
= 0.1-0.8 X_A^K + 1.4 (X_A^K)^2$ indicate the annihilation contributions.
The numerical coefficients are obtained for the input parameters discussed
before with $X_{H(A)}$ real, and using $X_{H(A)} = \ln(m_B/\Lambda)$. 
With these input parameters, 
$H_{\pi\pi}$ and $H_{K\pi}$ are in the range
between 0.8 to 1. 
This leads to very small annihilation
and hard scattering contributions as can be seen from 
Eq.~\ref{relgmod}. 
If one allows complex
values for $X_{H(A)}$, then the corrections can be larger \cite{3}, 
but we do not have a good estimate for these parameters. 

The most important SU(3) breaking effect that we have identified 
(in addition to the difference in decay constants) arises from
twist-2 distribution amplitudes. 
Using the central values
$\alpha^{\pi}_1 = 0$, $\alpha^{\pi}_2 =0.1$ and
$\alpha^K_1 = 0.3$, $\alpha_2^K = 0.1$,
the size of $\Delta^{\bar B^0}_{\pi^-\pi^+}/
\Delta^{\bar B^0}_{K^- \pi^+}$
increases by a factor of 1.3,
and is approximately $-0.87$. 
By taking into account the full range of values 
for the $\alpha$-parameters, the maximum and minimum
numerical values of the above ratios are, respectively, 
$-1.4$ and $-0.6$. This can be used to 
test the standard model and the QCD improved factorization. 
In particular, the sign of the rate difference is not changed 
by the SU(3) breaking effects we have considered. 

Similar calculations can be carried out for
$B_s \to PP$ decays. We find that the
ratio of differences
\begin{eqnarray}
{\Delta^{\bar B^0}_{\pi^-\pi^+}\over 
\Delta^{\bar B^0}_{K^- \pi^+}} \, \approx \,
{\Delta^{\bar B^0_s}_{\pi^- K^+}\over 
\Delta^{\bar B^0_s}_{K^- K^+}}
\label{indd}
\end{eqnarray}
is independent of the twist-2 distribution
amplitudes or meson decay constants.
This relation is particularly interesting
because it can be used to test
the SM with less uncertainties.
The related CP asymmetries will be expressed
in terms of the corresponding branching ratios
which are scaled by transition form factors as
\begin{eqnarray}
Br(\bar B^0 \to \pi^- \pi^+ ) & = & 
C \left({F^{B\to \pi}_0\over F^{B_s \to K}_0} \right)^2
Br(\bar B^0_s \to \pi^- K^+ ) 
{Ph^B_{\pi\pi}\over Ph^{B_s}_{\pi K}},
\nonumber\\
Br(\bar B^0 \to K ^- \pi^+) & = & 
C \left({F^{B\to \pi}_0\over F^{B_s \to K}_0} \right)^2 
Br(\bar B^0_s \to K^- K^+) 
{Ph^{B_d}_{\pi K}\over Ph^{B_s}_{KK}},
\end{eqnarray}
where $C=(m_B^2 \tau_{B_s} / m_{B_s}^2 \tau_B)$ and
$Ph^B_{P_1P_2} = [(1-(m_{P_1}+m_{P_2})^2/m_B^2)
(1-(m_{P_1}-m_{P_2})^2/m_B^2]^{1/2}/2m_B$
is the phase space factor. In order to test the SM,
one needs to know the form factors, these can be
obtained from other processes or from theoretical
calculations. Alternatively, one can use these relations to obtain
the ratio of the form factors using experimental data.

There are similar relations for $B\to PV$ decays\cite{7}. 
By replacing one of the final octet pseudoscalar mesons
with a corresponding octet vector meson in the previously
discussed cases, one obtains
\begin{eqnarray}
& &
\Delta^{\bar B^0}_{\rho^-\pi^+} = - 
\Delta^{\bar B_s^0}_{K^{*-} K^+} \approx 
\Delta^{\bar B^0_s}_{\rho^- K^+} = - 
\Delta^{\bar B^0}_{K^{*-} \pi^+} \,
\nonumber\\
& & 
\Delta^{\bar B^0}_{\pi^-\rho^+} = - 
\Delta^{\bar B_s^0}_{K^- K^{*+}} \approx 
\Delta^{\bar B^0_s}_{\pi^- K^{*+}} 
= - \Delta^{\bar B^0}_{K^{-} \rho^+} \, ,
\label{relspv}
\end{eqnarray}
where the approximate sign indicates relations that 
hold true only when annihilation contributions
are neglected. These relations, Eq.~\ref{relspv}, are again expected
to receive SU(3) breaking corrections. 
To estimate the SU(3) breaking effects we use 
the QCD improved factorization model once again. 

When $M_2$ (the meson which picks up the spectator)
is a vector meson, as for example in
$\bar B^0 \to \pi^- \rho^+$, $\bar B^0_s \to K^- K^{*+}$
and $\bar B^0 \to K^- \rho^+$, $\bar B_s^0 \to \pi^- K^{*+}$,
the corresponding decay amplitudes can be
obtained by replacing the form factor $F^{B\to P}_0$
with $A^{B\to V}_0$ and $r_{\chi}$ with $-r_\chi$ in
Eq.~\ref{abppimpfac}, and by using the same expressions 
for Eq.~\ref{facsbpp}, 
except for $H_{M_1M_2}$ which has no twist-3 terms. 
By neglecting the annihilation contributions, the
analogue of  Eq.~\ref{relgmod} is
\begin{equation}
\frac{\Delta^{\bar B^0}_{\pi^- 
\rho^+}}{\Delta^{\bar B^0_s}_{K^- 
K^{*+}}} \approx - \, {m_{B}\over m_{B_s}}
\frac{f_{\pi}^2}{f_K^2} \left( 
\frac{A_0^{B \to \rho}}{A_0^{B_s \to K^*}} \right)^2
\frac{1 + 110 \alpha_1^{\pi} + 15.5 \alpha_2^{\pi}}{1+ 
110 \alpha_1^K + 15.5 \alpha_2^K}\;, 
\label{relpvmod} 
\end{equation}
and the same for
$\Delta^{\bar B^0_s}_{\pi^- K^{*+}}/
\Delta^{\bar B^0}_{K^- \rho^+}$.
We observe the large coefficient of $\alpha_1$ in both 
the numerator and denominator of Eq.~\ref{relpvmod}. 
Since $\alpha_1^{\pi} =0$ and $\alpha_1^K = 0.3 \pm 0.3$,
the denominator has a vary large uncertainty, making a 
prediction for this asymmetry impossible within this framework. 
On the other hand, this provides an opportunity to constrain 
(or even to determine) $\alpha_1^K$ when the ratio in Eq.~\ref{relpvmod} 
is measured.

When $M_1$ is the vector meson, as in
$\bar{B}^0 \to \rho^- \pi^+$, $\bar{B}^0_s \to K^{*-} K^+$
and $\bar{B}^0_s \to \rho^- K^+$,
$\bar{B}^0 \to K^{*-} \pi^+$, the decay amplitudes can be 
obtained by replacing the $r_{\chi}$ factor in Eq.~\ref{abppimpfac} 
with $r_K^* = \frac{2 m_{K^*}}{m_b} \, 
\frac{f^{\perp}_{K^*}}{f_{K^*}} \approx 0.3$
(and similarly for $r_{\rho}$), and by removing the penguin
terms $P^{p,EW}_{2,3}$ in the expressions for
$a_6$ and $a_8$ in Eq.~\ref{facsbpp}, because the vector meson
is described only by a twist-2 distribution amplitude.
With all this we obtain:
\begin{equation}
\frac{\Delta^{\bar{B}^0}_{\rho^- \pi^+}}{\Delta^{
\bar{B}^0_s}_{K^{*-} K^+}} \approx - \, 
{m_{B}\over m_{B_s}}\frac{f_{\rho}^2}{f_{K^*}^2} \left( 
\frac{F_1^{B \to \pi}}{F_1^{B_s \to K}} \right)^2
\frac{1 - 1.25 \alpha_1^{\rho} - 0.18 \alpha_2^{\rho}}{1 
- 1.25 \alpha_1^{K^*} -0.18 \alpha_2^{K^*}} \;.
\label{relvpmod}
\end{equation}
Using the central values of the ranges
$\alpha_1^{\rho} =0$, $\alpha_2^{\rho}=0.16 \pm 0.09$,
$\alpha_1^{K^*}=0.18 \pm 0.05$,
$\alpha_2^{K^*}=0.05 \pm 0.05$\cite{8}
and taking $f_{\rho} \approx 0.96 f_{K^*}$\footnote{We extract this 
ratio $f_\rho/f_K \approx 0.96$ from 
$\Gamma(\tau^- \rightarrow \rho^- \nu)/\Gamma(\tau^- \rightarrow K^{*-}
\nu)$.}  we find:
\begin{equation}
\frac{\Delta^{\bar{B}^0}_{\rho^- \pi^+}}{
\Delta^{\bar{B}^0_s}_{K^{*-} K^+}}
\approx - \, 1.15 \left( \frac{F_1^{B 
\to \pi}}{F_1^{B_s \to K}} \right)^2
\; , \; \; \;
\frac{\Delta^{\bar{B}^0_s}_{\rho^- K^+}}{
\Delta^{\bar{B}^0}_{K^{*-} \pi^+}}
\approx - \, 1.15 \left( \frac{F_1^{B_s 
\to K}}{F_1^{B \to \pi}} \right)^2.
\end{equation}

Our calculations show that important SU(3)
breaking effects arise from the light-cone
distributions of mesons in addition to those already 
present in the decay constants.
These effects can only be estimated with large uncertainty 
because the parameters $\alpha^P_{1,2}$ are not
well determined at present. Using the currently allowed ranges 
we find,
\begin{equation}
A_{CP}(\pi^- \pi^+) \approx -
\left( 3.1^{+1.9}_{-0.9} \right) A_{CP}(K^- \pi^+ )\; ,
\end{equation}
which can also be used to test the standard model and the 
improved factorization model to some extent.

We have also shown that in the case of $B \to PV$,
when the pseudoscalar meson is factored out, SU(3)
breaking is large and estimates have very large uncertainty at present.
In the case when the vector meson is factored out, 
as in Eq.~\ref{relvpmod}, the corrections are smaller.

It is important to emphasize, however, that
there are relations which are independent of
$\alpha_{1,2}^i$ parameters and decay constants. Examples include 
Eq.~\ref{indd}, a corresponding relation for 
the ratio of branching ratios 
($Br(\bar B^0 \to \pi^- \pi^+)/Br(\bar B^0 \to K^- 
\pi^+) \approx Br(\bar B_s \to \pi^- K^+)/Br(\bar B_s 
\to K^- K^+)$), and their analogues in $B\to PV$ decays. 
These relations are more reliable than Eq.~\ref{relg} 
in the sense that they 
do not receive the main SU(3) breaking corrections that we 
have investigated. Although this observation relies on 
the QCD improved factorization model, it may be more robust 
than model predictions for absolute values of rates 
because it only involves ratios. 

A systematic framework to study SU(3) breaking in $B$ decays 
is, of course, needed before the relations we have presented 
can be used in precision tests of the standard model. 
With the estimates we have presented here, the relations 
are still useful. Large experimental violations of them would 
hint at possible new physics; at the very least they would provide 
information on the limitations of the QCD improved factorization 
model. To test some of the  relations that we have discussed, 
charmless hadronic two body $B_s$ decays must 
also be measured. This can not be done by the 
B-factories at present, but in the near future such $B_s$ decays 
will be studied at the Tevatron II and at LHCb.

\begin{flushleft}
{\bf\large Acknowledgments}
\end{flushleft}
This work was supported in part by US DOE
contract numbers DE-FG03-96ER40969, DE-FG02-01ER41155, the  
Taiwan NSC grant N91-2112-M-002-42
and in part by the Ministry of Education Academic
Excellence Project 89-N-FA01-1-4-3.
One of us, (M.A.D.) wishes to acknowledge the
kind hospitality of the University of Oregon
where this work was done.
She thanks the U.S. Department of State, the Council
for International Exchange of Scholars (C.I.E.S.)
and the Romanian-U.S. Fulbright Commission for
sponsoring her participation in the
Exchange Visitor Program no. G-1-0005.


\begin{thebibliography}{99}
\bibitem{1} Y. Karyotakis et al (BaBar Collaboration),
{\it BABAR explores CP violation}, talk at
ICHEP2002, Amsterdam, July 2002.
\bibitem{2} M. Yamuchi et al (Belle Collaboration),
{\it CP violation in $B$ mesons}, talk at
ICHEP2002, Amsterdam, July 2002.
\bibitem{3} M. Beneke, G. Buchalla, M. Neubert and
C.T. Sachrajda, Nucl. Phys. {\bf B606} (2001) 245.
\bibitem{4} N.G. Deshpande and X.G. He, Phys. Rev. Lett. 
{\bf 75}, (1995) 1703; X.G. He, Eur. Phys. J. {\bf C9} (1999) 443.
\bibitem{he} H.-K. Fu et al., e-print hep-ph/0206199.
\bibitem{5} D.S. Du, D.S. Yang and G.H. Zhu, Phys. Rev.
{\bf D64} (2001) 014036; M.Z. Yang and Y.D. Yang,
Nucl. Phys. {\bf B609} (2001) 469;
T. Muta, A. Sugamoto, M.Z. Yang and Y.D. Yang,
Phys. Rev. {\bf D62} (2000) 094020;
J. Sun, G.H. Zhu and D.S. Du, e-print hep-ph/0211154.
\bibitem{6} P. Ball, J. High Ener. Phys. {\bf 09}
(1998) 005.
\bibitem{7} N.G. Deshpande, X.G. He and J.Q. Shi,
Phys. Rev. {\bf D62} (2000) 034018.
\bibitem{8} P. Ball and V.M. Braun, hep-ph/9808229.
\end{thebibliography}
\end{document}